\documentclass{siamltex}\usepackage{url}  
\usepackage{euscript,amsmath,amssymb,amsfonts,graphicx,bm}
  \usepackage{epstopdf}
  \usepackage{graphics} 
 \usepackage[colorlinks=true]{hyperref}
 \hypersetup{urlcolor=blue, citecolor=red}
 \usepackage[latin1]{inputenc}

\newcommand{\x}{{\bf x}}

\newcommand{\e}{{\rm e}}

\newcommand{\R}{{\mathbb R}}
\renewcommand{\P}{{\mathbb P}}
\newcommand{\E}{{\mathbb E}}
\newcommand{\dis}{\displaystyle}
\newcommand{\calN}{{\mathcal N}}

\begin{document}
\title{Directional search-and-capture model of cytoneme-based morphogenesis\thanks{PCB
was supported by the National Science Foundation (DMS-1613048).}}

\author{Paul C. Bressloff\thanks{Department of Mathematics, University of Utah, Salt Lake City, UT 84112
USA ({\tt bressloff@math.utah.edu})}  }

\date{today}

\maketitle

\begin{abstract} In this paper we develop a directional search-and-capture model of cytoneme-based morphogenesis. We consider a single cytoneme nucleating from a source cell and searching for a set of $N$ target cells $\Omega_k\subset \R^d$, $k=1,\ldots,N$, with $d\geq 2$. We assume that each time the cytoneme nucleates, it grows in a random direction so that the probability of being oriented towards the $k$-th target is $p_k$ with $\sum_{k=1}^Np_k<1$. Hence, there is a non-zero probability of failure to find a target unless there is some mechanism for returning to the nucleation site and subsequently nucleating in a new direction. We model the latter as a one-dimensional search process with stochastic resetting, finite returns times and refractory periods. We use a renewal method to calculate the splitting probabilities and conditional mean first passage times (MFPTs) for the cytoneme to be captured by a given target cell. We then determine the steady-state accumulation of morphogen over the set of target cells following multiple rounds of search-and-capture events and morphogen degradation. This then yields the corresponding morphogen gradient across the set of target cells, whose steepness depends on the resetting rate. We illustrate the theory by considering a single layer of target cells, and discuss the extension to multiple cytonemes.

 \end{abstract}
%%%%%%%%%%%%%%%%%%%%%%%%%%%%%%%%%%%%%%%%%%%%%%%%%%%

 
\section{Introduction}
Cytonemes are thin, actin-rich filaments that can dynamically extend up to several hundred microns to form direct cell-to-cell contacts. There is increasing experimental evidence that these direct contacts allow the active transport of morphogen to embryonic cells during development \cite{Roy11,Gradilla13,Kornberg14,Stanganello16,Zhang19}. The precise biochemical and physical mechanisms underlying how cytonemes find their targets, form stable contacts and deliver their cargo to target cells are currently unknown. However, it has been hypothesized that cytonemes find their targets via a random search process based on alternating periods of retraction and growth \cite{Kornberg14}. Indeed, imaging studies in {\em Drosophila} \cite{Bischoff13} and chick \cite{Sanders13} show that cytonemes actively expand and contract. In the particular case of Wnt signaling in zebrafish
\cite{Stanganello16}, the morphogen Wnt is clustered at the membrane tip of a cytoneme, which nucleates from a source cell and dynamically grows until making contact with a target cell and delivering its cargo. However, a cytoneme can also switch to a shrinkage phase and rapidly retract (reset), which is analogous to a microtubule catastrophe \cite{Mitch84,Dogterom93}. 

It has been hypothesized that contact-mediated transport of morphogen by cytonemes provides an alternative to diffusion as a mechanism for setting up morphogen concentration gradients in embryonic tissue \cite{Kornberg14,Zhang19}. These gradients then instruct a spatial pattern of distinct cell differentiation pathways according to the local morphogen level.
The latter could depend on a number of factors. For example, in the case of Hedgehog (Hh) protein gradients in the {\em Drosophila} wing disc, these could include the lengths and number of contact points between pairs of cytonemes from different cells. On the other hand, Wnt signaling gradients in the zebrafish neural plate could depend on cytoneme lengths and the frequency of contacts by cytonemes. Another complicating factor is the expansion of the neural plate during development, which means that cells are continuously moving out of the
cytonemal area of influence \cite{Stanganello16}.

In contrast to diffusion-based morphogenesis, there have been a relatively small number of mathematical modeling studies of cytoneme-based morphogenesis \cite{Teimouri16,Bressloff18,Kim18,Bressloff19,Kim19}. These have focused on 1D models in which cytonemes grow and shrink in a fixed direction along a 1D array of  target cells. Transport occurs via two distinct mechanisms. The first involves active motor-driven transport of morphogen packets (vesicles) along static cytonemes with fixed contacts between a source cell and a target cell \cite{Teimouri16,Bressloff18,Kim18,Kim19}. The second, which is the one considered further here, is based on nucleating cytonemes from a source cell dynamically growing and shrinking until making contact with one of the partially absorbing target cells \cite{Bressloff19}. Morphogen is assumed to be localized at the tip of a growing cytoneme, which is delivered as a ``morphogen burst'' when the cytoneme makes temporary contact with the target cell before subsequently retracting. The delivery of a single burst can be analyzed in terms of a first passage time (FPT) problem with a sticky boundary at the source cell. The latter takes into account the exponentially distributed waiting time required for nucleation of a new growing filament, following any return to the source cell. After delivery of a morphogen burst, the cytoneme retracts and a new search-and-capture process is initiated. This then leads to a sequence of search-and-capture events, whereby morphogen accumulates in the target cells. Assuming that the build up of resources within each target is counterbalanced by degradation, there will exist a steady-state morphogen distribution in the long-time limit, which takes the form of a morphogen gradient. One way to calculate the statistics of resource accumulation is to formulate multiple search-and-capture events as a $G/M/\infty$ queue \cite{Bressloff19}.

Recently, we have shown how the search-and-capture model of cytoneme-based morphogenesis can be mapped onto a search process with stochastic resetting  \cite{Bressloff19a,Bressloff20c}. The latter type of process provides a general theoretical framework for understanding a wide range of naturally occurring search processes. The basic idea is that the position of a particle performing a stochastic search for some target is reset to a fixed location at a random sequence of times, which is typically (but not necessarily) generated by a Poisson process. In many cases there exists an optimal resetting rate for minimizing the mean first passage time (MFPT) to reach a target. This was originally established for Brownian motion in an unbounded domain  \cite{Evans11a,Evans11b,Evans14}, but has subsequently been observed in a wide range of stochastic processes (For a recent review see Ref. \cite{Evans20}.) Most models of stochastic resetting assume that resetting is instantaneous and that the search process is restarted immediately. However, in the particular application to cytoneme-based search-and-capture, retraction (resetting) of the cytoneme occurs at a finite speed, and once the cytoneme has returned to the source cell, there is a nucleation time before a new search process begins. In other words, one has to consider stochastic resetting with finite return times and refractory periods \cite{Evans19a,Pal19a,Pal20,Bressloff20c}. The advantage of formulating cytoneme dynamics in terms of a process with stochastic resetting is that one can apply various probabilistic methods such as renewal theory and conditional expectations.

There are a few cases where an effective 1D model is relevant. For example, in the {\em Drosophila} wing disc Hh is transported by cytonemes that are distributed along the basal side of the wing disc columnar epithelium \cite{Kornberg14}. In particular, Hh is moved from source cells in the posterior compartment to target cells in the anterior compartment, resulting in a morphogen gradient along the anterior-posterior axis. However, a more common geometric configuration is the transverse projection of cytonemes from source cells into a 2D or 3D region of target cells \cite{Zhang19}. This requires allowing cytonemes to search in different directions.

Therefore, in this paper we significantly extend our previous work by considering a directional search-and-capture model of cytoneme-based morphogenesis, which is introduced in \S 2. We consider a single cytoneme nucleating from a source cell and searching for a set of $N$ target cells $\Omega_k\subset \R^d$, $k=1,\ldots,N$, with $d\geq 2$. We assume that each time the cytoneme nucleates, it grows in a random direction so that the probability of being oriented towards the $k$-th target is $p_k$ with $\sum_{k=1}^Np_k<1$. Hence, there is a non-zero probability of failure to find a target unless there is some mechanism for returning to the nucleation site and subsequently nucleating in a new direction. We model the latter as a one-dimensional search process with stochastic resetting, finite returns times and refractory periods. In \S 3 we use a renewal method \cite{Pal19a,Pal20,Bressloff20c} to calculate the mean first passage time (MFPT) for the cytoneme to be captured by a single target cell, and determine its dependence on cytoneme length and the resetting rate. Allowing for the possibility of failure in the absence of resetting leads to the existence of an optimal resetting rate at which the MFPT is minimized. In \S 4 we consider the full multi-target problem. Again using renewal theory, we calculate the splitting probabilities and conditional MFPTs for capture by a target cell. In \S 5 we consider the steady-state distribution of morphogen over the set of target cells following multiple rounds of search-and-capture events combined with morphogen degradation. Finally, in \S 6 we illustrate the theory by considering the example of a single layer of target cells. We also indicate how to extend  the model to include multiple nucleation sites on the source cell or on a local cluster of source cells, provided that each nucleation site is independent and described by the same statistics.

\section{Directional search-and-capture model}

\begin{figure}[b!]
\centering
\includegraphics[width=7cm]{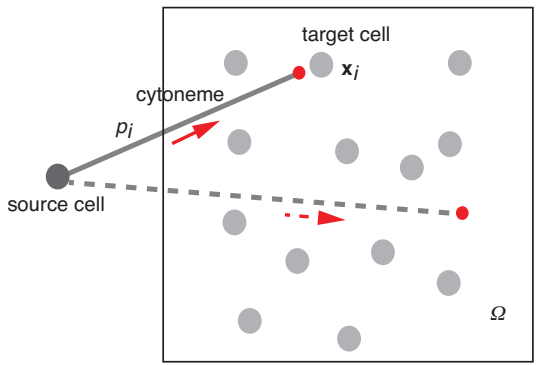} 
\caption{Schematic diagram of a single cytoneme nucleating from a source cell and penetrating a domain $\Omega\subset \R^d$ consisting of $N$ target cells labeled $i=1,\ldots,N$. The center of the $i$-th target cell is at $\x_i\in\Omega$. Each time a new cytoneme nucleates, it grows in a random direction such that the probability of being oriented towards the $i$-th target cell is $p_i$ with $p_{\rm tot}=\sum_{i=1}^Np_i<1$. The dashed line represents an orientation in which the cytoneme fails to find a target cell.}
\label{fig1}
\end{figure}

\begin{figure}[t!]
\centering
\includegraphics[width=8cm]{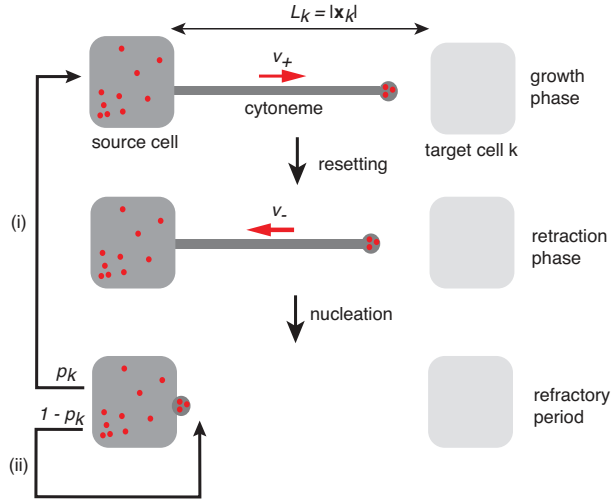} 
\caption{Search-and-capture model of cytoneme-based morphogen transport. A single cytoneme nucleates from a source cell and grows at a speed $v_+$. Each time it nucleates, the cytoneme is oriented towards the $k$-th target cell with probability $p_k$ (pathway (i)) or oriented away from the cell with probability $1-p_k$ (pathway (ii)). During a growth phase the cytoneme may randomly switch to a retraction phase (resetting) and return to the origin at speed $v_-$. After a refractory period (nucleation waiting time), a new cytoneme starts to grow and the process repeats. The distance of the target cell from the source cell is $L_k=|\x_k|$.}
\label{fig2}
\end{figure}

Consider a source cell with a single nucleation site from which a cytoneme grows towards one of $N$ target cells, $i=1,\ldots,N$, that are distributed in some bounded domain $\Omega\subset \R^d$, see Fig. \ref{fig1}. We will assume that the cytoneme nucleates in a random direction such that the probability of being oriented towards the $i$-th target cell at position $\x_i\in \Omega\subset \R^d$ is given by a fixed value $p_i$. (The source cell is taken to be at the origin.) The specific value of $p_i$ will depend on the angle subtended by the $i$-th target cell with respect to the source cell, which itself will depend on the cell size and its distance from the source cell. Given that there is a non-zero probability of the cytoneme searching in a ``wrong'' direction, that is, in a direction without a target, we take $p_{\rm tot}=\sum_{i=1}^Np_i<1$. Once a cytoneme has nucleated in a particular direction, it grows according to a dynamical process with stochastic resetting. Taking one end of the cytoneme to be fixed at $x=0$ (the nucleation site), the other end is represented by a stochastic variable $X(t)$, which can also be identified as the length of the cytoneme. Suppose that the cytoneme can exist in one of two discrete states: a right-moving (anterograde) state with speed $v_+$ or a left-moving (retrograde) state with speed $v_-$. The cytoneme undergoes the state transition $v_+\rightarrow v_-$ at a resetting rate $r$, after which it returns to the origin. At the origin the particle enters a refractory state for an exponentially distributed waiting time with rate $\eta$, prior to re-entering the anterograde state at a new, randomly selected orientation. The different stages of the search process for the $k$-th target cell is shown in Fig. \ref{fig2}.

\begin{figure}[t!]
\centering
\includegraphics[width=9cm]{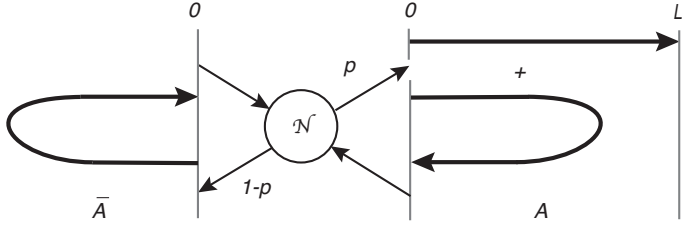}
\caption{Schematic illustration of different cytoneme states when $p<1$.}
\label{figs}
\end{figure}

Let us begin by considering a single target cell $(N=1$) at a distance $L$ from the source cell. Suppose that each time the cytoneme nucleates it grows towards the target cell with probability $p<1$ and grows in the wrong direction with probability $1-p$. (This is a simplified representation of the effects of other targets or failure to find a target, see \S 4.) In both cases, the cytoneme may switch to a shrinking phase at a resetting rate $r$ and return to the origin (nucleation site) at a speed $v_-$. It is convenient to partition the set of cytoneme states $\Sigma$ according to $\Sigma=\calN \cup A\cup  \overline{A}$,
where $\calN$ is the nucleation state, $A$ are the growing/shrinking states oriented in the direction of the target, and $\overline{A}$ are the corresponding states when the cytoneme is oriented in another direction. We denote the state of the cytoneme at time $t$ by $K(t)$. During each growth/shrinkage phase, let $p_n(x,t)$ be the probability density that at time $t$ the cytoneme tip is at a distance $X(t)=x$ from the nucleation site and in either the anterograde state ($n=+$) or the retrograde state ($n=-$). Similarly, let $P_0(t)$ denote the probability that the particle is in the refractory state at time $t$. If $K(t)\in A$ then $X(t)\in (0,L)$ evolves according to the Chapman-Kolmogorov (CK) equation
\begin{subequations} 
\label{B}
\begin{eqnarray}
	\frac{\partial p_+}{\partial t} &=& - v_+ \frac{\partial p_+}{\partial x} - r  p_+  \quad x\in (0,L), \\ 
	\frac{\partial p_-}{\partial t} &=& v_- \frac{\partial p_-}{\partial x} + r p_+,\quad \\
	\frac{dP_0}{dt}&=&v_-p_-(0,t)-\eta P_0(t),
\end{eqnarray}
together with the boundary conditions
\begin{equation}
v_+p_+(0,t)=\eta P_0(t),\quad p_-(L,t)=0.
\end{equation}
\end{subequations}
If the cytoneme hits $x=0$ ($x=L$) first then it enters the state $\calN$ (is captured by the target cell). On the other hand, if $K(t)\in \overline{A}$ then one simply waits a time $\sigma$ before the cytoneme returns to the state $\calN$. The waiting time $\sigma$ is a random variable with mean $\E[\sigma]=\bar{\sigma}$. We will assume that the cytoneme keeps growing until it resets so that
\begin{equation}
\bar{\sigma}=\frac{1}{r}\left (1+\frac{v_+}{v_-}\right ).
\end{equation}
(For simplicity, we ignore the possibility that the cytoneme hits another obstacle (not a target) and then retracts.) Finally, if $K(t)=\calN$ then the cytoneme transitions to a growing state, which either belongs to $A$ with probability $p$ or belongs to $\overline{A}$ with probability $1-p$. The time $\tau$ spent in state $\calN$ is exponentially distributed with mean time $\eta^{-1}$. A schematic diagram of the different states is shown in Fig. \ref{figs}. 

Note that the nucleation boundary condition at $x=0$ is mathematically identical to the so-called sticky boundary condition used in models of bidirectional transport and microtubular catastrophes \cite{Gopa11,Mulder12,Bressloff19}. The absorbing boundary condition at $x=L$ means that once the cytoneme reaches its target, it delivers its packet of morphogen and the search process ends.  We also impose the initial conditions
\[P_0(0)=0,\quad p_n(x,0)=\delta_{n,+}\delta(x).\]
That is, the cytoneme starts in the growth phase at $x=0$.
Summing equations (\ref{B}a) and (\ref{B}b) and then integrating with respect to $x$ over the interval $[0,L]$ shows that
\begin{eqnarray*}
 \frac{d}{dt}\int_0^Lp(x,t)dx&=-\left . [v_+p_+(x,t)-v_-p_-(x,t)]\right  |_0^L,
\end{eqnarray*}
where $p=p_++p_-$.
Given the boundary conditions (\ref{B}d), it follows that
\begin{equation*}
\frac{d}{dt}\int_0^Lp(x,t)dx+\frac{dP_0}{dt}=-v_+p_+(L,t)\equiv -J(t),
\end{equation*}
where $J(t)$ is the probability flux into the target.

\section{Mean first passage time for a single target}

Introducing the survival probability
\begin{equation}
\label{QQ}
Q(t)=\int_0^Lp(x,t)dx+P_0(t),
\end{equation}
we see that the first passage time density $f(t)$ is
\[f(t)=-\frac{dQ(t)}{dt}=J(t).\]
Hence, the mean first passage time $T$ for the cytoneme to be captured by the target when $K(t)=A$, assuming that it starts at $x=0$ is given by
\begin{equation}
T=-\int_0^{\infty}t\frac{dQ(t)}{dt}dt=\int_0^{\infty}Q(t)dt.
\end{equation}
We now calculate the MFPT to find the target in the two cases $p=1$ and $p<1$.

\subsection{Case $p=1$} 
\begin{figure}[t!]
\centering
\includegraphics[width=10cm]{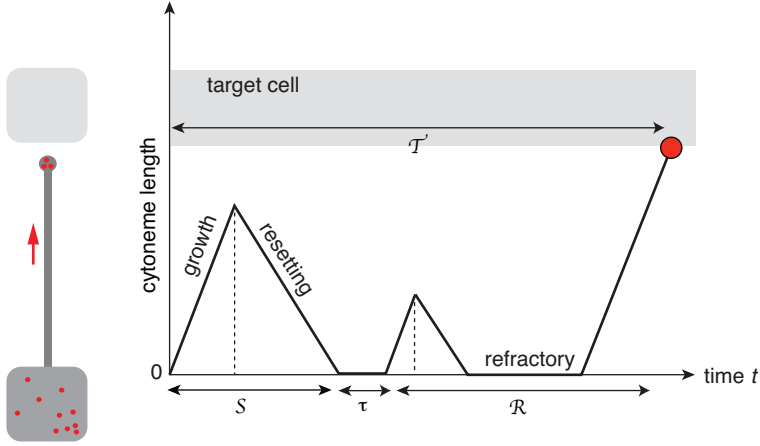} 
\caption{Mapping of cytoneme search-and capture to a first passage time problem for a particle under stochastic resetting. Resetting events are Poissonian with a rate $r$. The nucleation intervals are generated by an exponential waiting time density $\psi(\tau)=\eta \e^{-\eta \tau}$ with rate $\eta$. There is an absorbing boundary at $x=L$. Also shown is the decomposition of the first passage time according to the sum ${\mathcal T}={\mathcal S}+\tau+{\mathcal R}$.}
\label{fig4}
\end{figure}

As we have recently highlighted elsewhere \cite{Bressloff20c}, when $p=1$ (zero probability of failure) the above model can be mapped onto a stochastic resetting process for a particle with a refractory (nucleation) period \cite{Evans19a} and a finite return time \cite{Pal19a,Pal20}. This is illustrated in Fig.   \ref{fig4}, which shows a sample particle trajectory prior to capture by the target at $x=L$. We can thus use renewal theory to determine the mean first passage time $T$ in terms of the Laplace transform of the survival probability without reset, which we denote by $Q_0(t)$. For the given system, the latter is defined according to
$Q_0(t)=\int_0^Lp_+(x,t)dx$,
with
\begin{eqnarray}
\label{Bp}
	\frac{\partial p_+}{\partial t} &=& - v_+ \frac{\partial p_+}{\partial x}   \quad x\in (0,L).
\end{eqnarray}
This has the solution $p_+(x,t)=\delta(x-v_+t)$, which means that $Q_0(t)=H(L/v_+-t)$ where $H$ is the Heaviside function. Moreover, the Laplace transform is
\begin{equation}
\label{Q0}
\widetilde{Q}_0(s)=\int_0^{\infty}\e^{-st}H(L/v_+-t)dt=\int_0^{L/v_+}\e^{-st}dt=\frac{1}{s}\left (1-\e^{-sL/v_+}\right ).
\end{equation}

Here we use the particular formulation introduced in Ref. \cite{Bressloff20c}. 
This exploits the fact that resetting eliminates any memory of previous search stages. Consider the following set of first passage times;
 \begin{eqnarray}
	\mathcal{T} &=& \inf \{ t>0;   X(t)=L\}, \nonumber \\
	\mathcal{S} &=& \inf \{ t>0; X(t) =0\}, 
	\label{FPTa} \\
	\mathcal{R} &=& \inf \{ t>0;   X(t+{\mathcal S}+\tau)=L\}.\nonumber
\end{eqnarray}
Here ${\mathcal T}$ is the FPT for finding the target irrespective of the number of resettings, ${\mathcal S}$ is the FPT for the first resetting and return to the origin, $\tau$ is the first refractory period, and ${\mathcal R}$ is the FPT for finding the target given that at least one resetting has occurred. Next we introduce the sets $\Omega = \{ \mathcal{T}< \infty \}$ and $\Gamma= \{ \mathcal{S} < \mathcal{T}< \infty\}\subset \Omega$.
That is, $\Omega$ is the set of all events for which the particle is eventually absorbed by the target (which has measure one), and $\Gamma$ is the subset of events in $\Omega$ for which the particle resets at least once. It immediately follows that 
$\Omega \backslash \Gamma = \{ \mathcal{T} < {\mathcal S} = \infty \}$.
In other words, $\Omega\backslash \Gamma$ is the set of all events for which the particle is captured by the target without any resetting. We now use a probabilistic argument to calculate the MFPT $T=\E[{\mathcal T}]$ in the presence of resetting ($r>0$).

Consider the decomposition
\begin{eqnarray}
	\E[{\mathcal T}]& =& \mathbb{E}[\mathcal{T}1_{\Omega \backslash \Gamma}]+ \mathbb{E}[\mathcal{T}1_{\Gamma}].
\end{eqnarray}
The first expectation on the right-hand side can be evaluated by noting that it is the MFPT for capture by the target without any resetting, and the probability density for such an event is $-\e^{-rt}\partial_tQ_0(t)$. Hence,
\begin{equation}
\label{E10}
 \mathbb{E}[\mathcal{T}1_{\Omega\backslash \Gamma}]=-\int_0^{\infty}t\e^{-rt}\frac{dQ_0(t)}{d t}dt=\left (1+r\frac{d}{dr}\right )\widetilde{Q}_0(r),
 \end{equation}
 where $\widetilde{Q}_0(r)$ is the Laplace transform of $Q_0(t)$ with $s=r$.
The second expectation can be further decomposed as
\begin{eqnarray}
  \mathbb{E}[\mathcal{T}1_{\Gamma}] 
	&=&\mathbb{E}[({\mathcal S}+\tau+{\mathcal R})1_{\Gamma}]=\mathbb{E}[{\mathcal S}1_{\Gamma}]+\overline{\tau} \P[\Gamma]+\mathbb{E}[{\mathcal R}1_{\Gamma}\nonumber ]\\
	&=&\mathbb{E}[{\mathcal S}1_{\Gamma}]+(\overline{\tau}+T) \P[\Gamma].
	\label{E20}
	\end{eqnarray}  
Here $\E[\tau]=\overline{\tau}=\eta^{-1}$ is the mean refractory period, and we have used the result
$\mathbb{E}[{\mathcal R}1_{\Gamma}]=TP[\Gamma]$. The latter follows from the fact that return to the origin restarts the stochastic process without any memory.

In order to calculate $\mathbb{E}[{\mathcal S}1_{\Gamma}]$, it is necessary to incorporate the time to return to the origin following the first return event. The first resetting occurs with probability 
$r\e^{-rt}Q_0(t)dt$ in the interval $[t,t+dt]$. At time $t$ the particle is at position $v_+t$ and thus takes an additional time $v_+t/v_-$ to return to $x=0$. We thus find
\begin{eqnarray}
 \mathbb{E}[{\mathcal S}1_{\Gamma}]&=\int_0^{\infty} r  \e^{-rt} t\left (1+\frac{v_+}{v_-}\right )Q_0(t)dt=-r\left (1+\frac{v_+}{v_-}\right )\frac{d}{dr}\widetilde{Q}_0(r).
\label{E30}
\end{eqnarray}	
Moreover, from the definitions of the first passage times and the effect of resetting, 
\begin{equation}
 \P[\Gamma]=\P[{\mathcal S}<\infty]\P[{\mathcal R}<\infty],
\label{E40}
\end{equation}
with $\P[{\mathcal R}<\infty]=1$ and
\begin{eqnarray}
 \P[{\mathcal S}<\infty]&=\int_0^{\infty}r  \e^{-rt} Q_0(t)dt=r\widetilde{Q}_0(r).
\label{E60}
\end{eqnarray}
Combining equations (\ref{E10})--(\ref{E60}) yields the implicit equation
\begin{eqnarray}
 T=\left (1+r\frac{d}{dr}\right )\widetilde{Q}_0(r)+r\overline{\tau}\widetilde{Q}_0(r)-r \left (1+\frac{v_+}{v_-}\right )\frac{d}{dr}\widetilde{Q}_0(r)+r\widetilde{Q}_0(r)T.
\end{eqnarray}
Rearranging this equation yields the result 
\begin{equation}
\label{Tr2}
T=\frac{\widetilde{Q}_0(r)+r\overline{\tau}\widetilde{Q}_0(r)-r\frac{\dis v_+}{\dis v_-}\widetilde{Q}_0'(r),}{1-r\widetilde{Q}_0(r)}.
\end{equation}
Note that this is a general formula for a dynamical process with stochastic resetting, finite return times and refractory periods \cite{Pal19a,Pal20,Bressloff20c}. For our particular model, we
substitute equation (\ref{Q0}) into equation (\ref{Tr2}) to give
\begin{equation}
T=T(L):=\frac{1}{r}\left [(\e^{rL/v_+}-1)(1+r\overline{\tau}+v_+/v_-) -\frac{\dis rL}{\dis v_-} \right ].
\end{equation}
In the limit $r\rightarrow 0$, $T(L)\rightarrow L/v_+$, which is simply the deterministic time for the cytoneme tip to travel the distance $L$.

\subsection{Case $p<1$}

 In order to include the effects of failure in the absence of resetting, it is necessary to generalize the FPTs defined by equations (\ref{FPTa}).  Let $I(t)$ denote the number of resettings in the interval $[0,t]$. Assuming that the cytoneme starts out in the growing state, we set
\begin{eqnarray}
{\mathcal T}&=&\inf\{t\geq 0; X(t)=L,\ I(t)\geq 0\},\nonumber \\
{\mathcal T}_A &=&\inf\{t\geq 0; X(t)=L,\ I(t)\geq 0 |K(0)\in A\},\nonumber \\
\label{st}
{\mathcal S}_A &=&\inf\{t\geq 0; X(t)=0, \ I(t)=1|K(0)\in A\},\\
{\mathcal R}_{\overline{A}}&=&\inf\{t\geq 0; X(t+\tau+\sigma)=L, \ I(t+\tau+\sigma)\geq 1|K(0)\in \overline{A}\},\nonumber \\
{\mathcal R}_A &=&\inf\{t\geq 0; X(t+\tau+{\mathcal S}_A)=L,\ I(t+\tau+{\mathcal S}_A)\geq 1|K(0)\in A\}.\nonumber
\end{eqnarray}
These are the natural extensions of the FPTs defined in equation (\ref{FPTa}), which keep track of whether or not the nucleating cytoneme is oriented towards the given target. Next we define the sets
\begin{align}
\Omega &= \{ \mathcal{T}< \infty \}, \quad \Omega_{A}= \{  \mathcal{T} < \infty\}\cap \{K(0)=A\}\subset \Omega,\nonumber \\
\Gamma&=  \{ \mathcal{S}_A < \mathcal{T}_A < \infty\}\subset \Omega_{A},\quad  \overline{\Gamma}= \{  \mathcal{T} < \infty\}\cap \{K(0)=\overline{A}\}\subset \Omega
\end{align}
where $\Omega$ is the set of all events for which the cytoneme is eventually absorbed by the target ($\P[\Omega]=1$), $\Omega_A$ is the subset of events conditioned on starting in the state $A$, and $\Gamma$ ($\overline{\Gamma}$) is the subset of events in $\Omega_A$ ($\Omega$) conditioned on starting in the state $A$ ($\overline{A}$). It follows that $\Omega=\Omega_{A}\cup \overline{\Gamma}$, and
\[
	\Omega_{A} \backslash \Gamma= \{ \mathcal{T} < {\mathcal S}_A = \infty \},
\]
where $\Omega_{A}\backslash \Gamma$ is the set of all events for which the cytoneme is captured by the target without any resettings.

For $p <1$ we have
\begin{align}
 T_p:=\E[{\mathcal T}]&=p\E[{\mathcal T}_A1_{\Omega_A}]+(1-p) \E[({\mathcal R}_{\overline{A}}+\tau+\sigma)1_{\overline{\Gamma}}]\nonumber 
 \\ &=p\E[{\mathcal T}_A1_{\Omega_A}]+(1-p) (\bar{\tau}+\bar{\sigma}+\E[{\mathcal R}_{\overline{A}}1_{\overline{\Gamma}}])\nonumber \\
 &=(1-p)(\bar{\sigma}+\bar{\tau})+p\E[{\mathcal T}_A1_{\Omega_A}]+(1-p)\E[{\mathcal T}],
 \label{GGtau}
\end{align}
where $\bar{\sigma}=r^{-1}(1+v_+/v_-)$, and we have again used the fact that return to the origin restarts the stochastic process without any memory so $\E[{\mathcal R}_{\overline{A}}]=\E[{\mathcal T}]$. Rearranging, 
\begin{equation}
\label{EE1}
T_p=\frac{(1-p)[\bar{\tau}+\bar{\sigma}]}{p}+\E[{\mathcal T}_A1_{\Omega_A}].
\end{equation}
The analysis of $\E[{\mathcal T}_A1_{\Omega_A}]$ proceeds along similar lines to the case $p=1$ by performing the decomposition
\begin{eqnarray}
\E[{\mathcal T}_A1_{\Omega_A}]&=&\E[{\mathcal T}_A1_{\Omega_A\backslash\Gamma}] +\E[{\mathcal T}_A1_{\Gamma}]=\E[{\mathcal T}_A1_{\Omega_A\backslash\Gamma}]+\E[({\mathcal S}_A+\tau +{\mathcal R})1_{\Gamma}]\nonumber \\
&=&\E[{\mathcal T}_A1_{\Omega_A\backslash\Gamma}]+\E[{\mathcal S}_A1_{\Gamma}]+(\bar{\tau}+T_p)\P[\Gamma],
\label{EE2}
\end{eqnarray}
with $P[\Gamma]=r\widetilde{Q}_0(r)$, see equations (\ref{E40}) and (\ref{E60}).
The term $ \mathbb{E}[\mathcal{T}_A1_{\Omega_A\backslash \Gamma}]$ is the MFPT for capture by the target without any resetting, given that the cytoneme is oriented in the correct direction, and is thus given by equation (\ref{E10}): 
\begin{equation}
\label{EE4}
 \mathbb{E}[\mathcal{T}_A1_{\Omega_A\backslash \Gamma}]=\left (1+r\frac{d}{dr}\right )\widetilde{Q}_0(r).
 \end{equation}
Similarly, $\mathbb{E}[{\mathcal S}_A1_{\Gamma}]$ is given by equation (\ref{E30}):
\begin{eqnarray}
\label{EE5}
 \mathbb{E}[{\mathcal S}_A1_{\Gamma}]&=-r\left (1+\frac{v_+}{v_-}\right )\frac{d}{dr}\widetilde{Q}_0(r).
\end{eqnarray}	
Finally, combining equations (\ref{EE1})--(\ref{EE5}) yields the implicit equation
\begin{eqnarray}
 T_p=\frac{(1-p)[\bar{\tau}+\bar{\sigma}]}{p}+\left (1+r\frac{d}{dr}\right )\widetilde{Q}_0(r)+r\widetilde{Q}_0(r)(\overline{\tau}+T_p))-r \left (1+\frac{v_+}{v_-}\right )\frac{d}{dr}\widetilde{Q}_0(r).\nonumber\\
\end{eqnarray}
Rearranging this equation yields the result 
\begin{equation}
T_p=\frac{1}{1-r\widetilde{Q}_0(r)}\left \{\frac{(1-p)[\bar{\tau}+\bar{\sigma}]}{p}+\widetilde{Q}_0(r)+r\overline{\tau}\widetilde{Q}_0(r)-r\frac{\dis v_+}{\dis v_-}\widetilde{Q}_0'(r)\right \}.
\label{Tp}
\end{equation}
The first term in $\{\cdot\}$ has a simple interpretation, as can be seen by noting that
\begin{align*}
\frac{(1-p)[\bar{\tau}+\bar{\sigma}]}{p}&=\frac{(1-p)[\bar{\tau}+\bar{\sigma}]}{1-(1-p)}=[\bar{\tau}+\bar{\sigma}]\sum_{m=1}^{\infty}(1-p)^m.
\end{align*}
That is, each time there is an excursion in the wrong direction, which occurs with probability $(1-p)$, an additional mean time penalty of $\bar{\tau}+\bar{\sigma}$ is incurred.
Finally, substituting for $\widetilde{Q}_0(r)$ using equation (\ref{Q0}) and setting $\bar{\sigma}=r^{-1}(1+v_+/v_-)$, we obtain the result
\begin{equation}
\label{Tp2}
T_p=T_p(L):=\frac{(1-p)[\bar{\tau}+r^{-1}(1+v_+/v_-)]}{p}\e^{rL/v_+}+T(L),
\end{equation}
with $T(L)$ given by equation (\ref{Tr2}). 

\begin{figure}[t!]
\centering
\includegraphics[width=12cm]{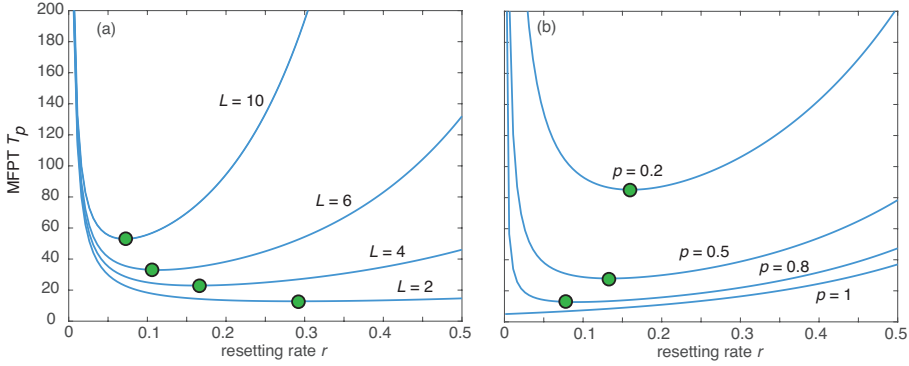}
\caption{Plots of MFPT $T_p(L)$ as a function of the resetting rate $r$. (a) Various cytoneme lengths $L$ and $p=0.5$. (b) Various probabilities $p$ and $L=5$. Dimensionless units with $v_+=1,\bar{\tau}=1,v_-=5$. Optimal resetting rates are indicated by the filled circles.}
\label{fig5}
\end{figure}

It is clear from equation (\ref{Tp2}) that $T_p(L)\rightarrow \infty$ as $r\rightarrow 0$, since $\bar{\sigma}\rightarrow \infty$, which is a consequence of the possible failure of the search process in the absence of resetting. This suggests that there exists an optimal resetting rate at which the MFPT is minimized, which is indeed found to be the case. In Fig. \ref{fig5} we show plots of $T_p$ as a function of the resetting rate for various distances $L$ and probabilities $p$. It can be seen that the optimal resetting rate $r_{\rm opt}$ at which each curve has a minimum is an increasing function of $L$ and a decreasing function of $p$, with $r_{\rm opt}\rightarrow 0$ as $p\rightarrow 1$.

 \section{Splitting probabilities and conditional MFPTs for multiple targets} 
 
We now extend the analysis of \S 3 in order to calculate the splitting probability $\pi_j$ and conditional MFPT $T_j$ to be captured by the $j$-th target cell, $j=1,\ldots,N$, without previously being captured by any other target. The set of cytoneme states is partitioned as $\Sigma=\calN\cup \overline{A}\cup_{j=1}^N A_j$,
where $\calN$ is the nucleation state, $A_j$ are the growing/shrinking states oriented in the direction of the $j$-th target for $j=1,\ldots,N$, and $\overline{A}$ is any state oriented away from all of the targets. Let $K(t)$ denote the state of the cytoneme at time $t$. If $K(t)\in A_j$ then $X(t)\in (0,L_j)$ evolves according to equations (\ref{B}) with $L=L_j$. If the cytoneme hits $x=0$ first then it enters the state $\calN$, otherwise it is captured by the $j$-th target cell. If $K(t)=\calN$ then the cytoneme transitions to a new growing state in one of the states $A_k$ with probability $p_k$ or the state $\overline{A}$ with probability $\bar{p}=1-\sum_{i=1}^Np_i$, after a waiting time $\tau$. Again let $I(t)$ denote the number of resettings in the interval $[0,t]$. Assuming that the cytoneme starts out
in the growing state, we introduce a set of FPTs, which are the multi-target analogs of equation (\ref{st}):
\begin{eqnarray}
{\mathcal T}_{j}&=&\inf\{t\geq 0; X(t )=L_j, \ I(t)\geq 0\},\nonumber \\
\widehat{\mathcal T}_{j}&=&\inf\{t\geq 0; X(t )=L_j, \ I(t)\geq 0|K(0)=A_j\},\nonumber \\
\label{stm}
{\mathcal S}_j&=&\inf\{t\geq 0; X(t )=0,\ I(t)=1|K(0)\in A_j  \},\\
{\mathcal R}_{ji}&=&\inf\{t\geq 0; X(t+\tau+{\mathcal S}_i )=L_j,\ I(t+\tau+{\mathcal S}_i )\geq 1|K(0)\in A_i\},\nonumber \\
\overline{\mathcal R}_{j}&=&\inf\{t\geq 0; X(t +\tau+\sigma)=L_j, \ I(t +\tau+\sigma)\geq 1|K(0)\in \overline{A}\},\nonumber
\end{eqnarray}
Here ${\mathcal T}_j$ is the FPT for finding the $j$-th target irrespective of the number of resettings, $\widehat{\mathcal T}_j$ is the corresponding FPT conditioned on starting in the state $A_j$, ${\mathcal S}_j$ is the FPT for the first resetting and return to the origin starting from the state $A_j$, ${\mathcal R}_{ji}$ is the FPT for finding the $j$-th target after at least one resetting conditioned on starting from the state $A_i$, and $\overline{\mathcal R}_j$ is the analogous FPT starting from the failure state $\overline{A}$. Next we define the sets
\begin{align}
\Omega_j &= \{ \mathcal{T}_j< \infty \}, \quad \Omega_{ji} = \{  \mathcal{T}_j < \infty\}\cap \{K(0)=A_i\}\subset \Omega_j,\nonumber \\
\Gamma_j&=  \{ \mathcal{S}_j < \mathcal{T}_j < \infty\}\subset \Omega_{jj},\quad  \overline{\Gamma}_{j}= \{  \mathcal{T}_j < \infty\}\cap \{K(0)=\overline{A}\}\subset \Omega_j
\end{align}
where $\Omega_j$ is the set of all events for which the cytoneme is eventually absorbed by the $j$-th target cell without being absorbed by any other target, $\Omega_{ji}$ ($\overline{\Gamma}_j$) is the subset of events in $\Omega_j$ conditioned on starting in the state $A_i$ ($\overline{A}$), and $\Gamma_j$ is the subset of events in $\Omega_{jj}$ that reset at least once. It follows that $\Omega_j=
\cup_{i=1}^N\Omega_{ji}\cup \overline{\Gamma}_j$, and
\[
	\Omega_{jj} \backslash \Gamma_{j}= \{ \mathcal{T} _j< {\mathcal S}_j = \infty \},
\]
where $\Omega_{jj}\backslash \Gamma_{j}$ is the set of all events for which the cytoneme is captured by the $j$-th target without any resettings. During each search phase directed towards the $j$-th target, we denote the survival probability without resetting by $Q_j(t)$, whose Laplace transform is
\begin{equation}
\label{Qj}
\widetilde{Q}_j(s)=\frac{1}{r}\left (1-\e^{-sL_j/v_+}\right ).
\end{equation}

The splitting probability $\pi_j$ can be decomposed as 
\begin{equation} \label{split0}
\pi_j:= \mathbb{P}[\Omega_j] =  p_j\mathbb{P}[\Omega_{jj}] +\sum_{i\neq j}p_i  \mathbb{P}[\Omega_{ji}]+\overline{p}\mathbb{P}[\overline{\Gamma}_{j}],
\end{equation}
We have the further decomposition
\begin{equation}
\label{split1}
\mathbb{P}[\Omega_{jj}] =\mathbb{P}[\Omega_{jj}\backslash \Gamma_j] +\mathbb{P}[\Gamma_{j}] 
\end{equation}
Let us consider the latter decomposition first. The probability that the cytoneme is captured by the $j$-th target in the interval $[t,t+dt]$ without any resettings is 
\begin{subequations}
\begin{eqnarray}
\P[\Omega_{jj}\backslash\Gamma_j]&= -\int_0^{\infty}\e^{-rt}\frac{dQ_{j}(t)}{dt} dt =\left (1-r\widetilde{Q}_j(r)\right ).
\label{PS0}
\end{eqnarray}
Next, from the definitions of the first passage times, we have
\begin{equation}
\label{PS}
\mathbb{P}[\Gamma_j]=\P[{\mathcal S}_j<\infty]\P[{\mathcal R}_{jj}<\infty]=\P[{\mathcal S}_j<\infty]\pi_j.
\end{equation}
We have used the renewal property of resetting to set $\P[{\mathcal R}_{jj}<\infty]=\pi_j$. The probability $\P[{\mathcal S}_j<\infty]$ is determined by noting that during a growth phase in the $j$-th direction, we require that the cytoneme returns to the origin before reaching the target at $L_j$. The probability of first switching to the return phase in the time interval $[t,t + dt]$ is equal to the product of the reset probability $r\e^{-rt}dt$ and the survival probability $Q_j(t)$. Hence,
\begin{eqnarray}
 \P[{\mathcal S}_j<\infty]&=&\int_0^{\infty}r  \e^{-rt} Q_j(t)dt=r\widetilde{Q}_j(r).
\label{PS1}
\end{eqnarray}
Finally, turning to the decomposition (\ref{split0}), we have
\begin{equation}
\label{PS2}
\mathbb{P}[\Omega_{ji}]=\P[{\mathcal S}_i<\infty]\P[{\mathcal R}_{ji}<\infty] =
r\widetilde{Q}_i(r)\pi_j,
\end{equation}
and
\begin{equation}
\label{PS3}
\mathbb{P}[\overline{\Gamma}_{j}]=\P[\overline{\mathcal R}_j<\infty] =\pi_j
\end{equation}
\end{subequations}

Combining equations (\ref{PS0})--(\ref{PS3}) yields the implicit equation 
\[\pi_{j} =p_j\left (1-r\widetilde{Q}_j(r)\right )+r\pi_{j}\sum_{l=1}^N p_l\widetilde{Q}_{l}(r)+\overline{p}\pi_j,\]
which on rearranging gives
\begin{equation}
\label{Pie}
\pi_{j}=\frac{p_j\left (1-r\widetilde{Q}_j(r)\right )}{ 1-
r\sum_{l=1}^Np_l \widetilde{Q}_{l}(r)-\bar{p}}=\frac{p_j\left (1-r\widetilde{Q}_j(r)\right )}{ \sum_{l=1}^Np_l\left (1-r\widetilde{Q}_l(r)\right )}.
\end{equation}
Summing both sides of equation (\ref{Pie}) with respect to $j$ implies that 
\begin{equation}
\sum_{j=1}^N\pi_j = 1.
\end{equation}
That is, in the presence of resetting, the probability of eventually finding a target is unity.

Similarly, we decompose the MFPT $\E[{\mathcal T}_j1_{\Omega_j}]:=\pi_jT_j$ as
\begin{eqnarray}
	\E[{\mathcal T}_j1_{\Omega_j}]& =&p_j \mathbb{E}[\widehat{\mathcal{T}}_j1_{\Omega_{jj} \backslash \Gamma_{j}}]+ p_j\mathbb{E}[\widehat{\mathcal{T}}_j1_{\Gamma_{j}}]+
	\sum_{i\neq j}p_i\mathbb{E}[\mathcal{T}_j1_{\Omega_{ji}}]+\overline{p}\mathbb{E}[\mathcal{T}_j1_{\overline{\Gamma}_{j}}].
\end{eqnarray}
The first expectation can be evaluated by noting that it is the MFPT for capture by the $j$-th target without any resetting. From equation (\ref{E10}), we thus have
\begin{subequations}
\begin{equation}
\label{E1}
 \mathbb{E}[\widehat{\mathcal{T}}_j1_{\Omega_{jj}\backslash \Gamma_{j}}]=\left [1+r\frac{d}{dr}\right ]\widetilde{Q}_{j}(r).
 \end{equation}
The second expectation is further decomposed as
\begin{eqnarray}
\label{E2}
  \mathbb{E}[\widehat{\mathcal{T}}_j1_{\Gamma_j}] 
	&=\mathbb{E}[({\mathcal S}_j+ {\tau}+{\mathcal R}_{jj})1_{\Gamma_j}]=\mathbb{E}[{\mathcal S}_j1_{\Gamma_j}]+\bar{\tau}\P[\Gamma_j]+\mathbb{E}[{\mathcal R}_{jj}1_{\Gamma_j}].
	\end{eqnarray} 
In order to calculate $\mathbb{E}[{\mathcal S}_j1_{\Gamma_j}]$, we need to calculate the mean time that the cytoneme returns to the origin before reaching the target at $L_j$. Following along analogous lines to equation (\ref{E30}), we have
\begin{eqnarray}
 \mathbb{E}[{\mathcal S}_j1_{\Gamma_j}]&=&\P[{\mathcal R}_{jj}<\infty] \int_0^{\infty} r  \e^{-rt} t \left (1+\frac{v_+}{v_-}\right )Q_j(t)dt\nonumber \\
 &=&-r\pi_j\left (1+\frac{v_+}{v_-}\right ) \frac{d}{dr}\widetilde{Q}_j(r).
\end{eqnarray}	
Hence,
\begin{align}
\label{E3}
  \mathbb{E}[\widehat{\mathcal{T}}_j1_{\Gamma_j}] 
	&=-r\pi_j\left (1+\frac{v_+}{v_-}\right ) \frac{d}{dr}\widetilde{Q}_j(r)+r\widetilde{Q}_j(r)\pi_j(\bar{\tau}+T_j).
	\end{align}
In addition,
\begin{align}
\label{E4}
\mathbb{E}[\mathcal{T}_j1_{\Omega_{ji}}]&=\mathbb{E}[({\mathcal S}_i+ {\tau}+{\mathcal R}_{ji})1_{\Omega_{ji}}]=\mathbb{E}[{\mathcal S}_i1_{\Omega_{ji}}]+\bar{\tau}\P[{\Omega_{ji}}]+\mathbb{E}[{\mathcal R}_{jj}1_{\Omega_{ji}}]\\
&=-r\pi_i\left (1+\frac{v_+}{v_-}\right ) \frac{d}{dr}\widetilde{Q}_i(r)+r\widetilde{Q}_i(r)\pi_j(\bar{\tau}+T_i).
\nonumber
\end{align}
Finally, 
\begin{align}
\label{E5}
\mathbb{E}[\mathcal{T}_j1_{\overline{\Gamma}_{j}}]&=\mathbb{E}[(\sigma+ {\tau}+\overline{\mathcal R}_{j})1_{\overline{\Gamma}_{j}}]=(\bar{\sigma}+\bar{\tau})\P[{\overline{\Gamma}_{j}}]+\mathbb{E}[{\overline{\mathcal R}}_{j}1_{\overline{\Gamma}_{j}}]\\
&=\pi_j(\bar{\sigma}+\bar{\tau}+T_j).\nonumber
\end{align}
\end{subequations}

Combining equations (\ref{E1})--(\ref{E5}) yields the implicit equation
\begin{eqnarray}
	\pi_j T_j& =& p_j\left [1+r\frac{d}{dr}\right ]\widetilde{Q}_{j}(r)-r\pi_j\left (1+\frac{v_+}{v_-}\right )\left [\sum_{l=1}^Np_l \frac{d}{dr}\widetilde{Q}_l(r)\right ]\nonumber\\
	&&\quad +r\left (\bar{\tau}+T_j\right )\pi_j\left [\sum_{l=1}^Np_l\widetilde{Q}_{l}(r)\right ]+\bar{p}\pi_j(\bar{\sigma}+\bar{\tau}+T_j),
\end{eqnarray}
which can be arranged to give the general result
\begin{eqnarray}
\label{Tk}
\pi_jT_j& =&\left \{p_j\left [1+r\frac{d}{dr}\right ]\widetilde{Q}_{j}(r)+r\pi_j\sum_{l=1}^Np_l \left [\bar{\tau}\widetilde{Q}_{l}(r)-\left (1+\frac{v_+}{v_-}\right ) \frac{d}{dr}\widetilde{Q}_l(r) \right ]+\bar{p}\pi_j(\bar{\sigma}+\bar{\tau})\right \}\nonumber \\
&&\quad \times \left \{\frac{1}{1-r\sum_{l=1}^Np_l\widetilde{Q}_{l}(r)-\bar{p}}\right \}.
\end{eqnarray}
Substituting for $\widetilde{Q}_{j}(r)$ using equation (\ref{Qj}), equations (\ref{Pie}) and (\ref{Tk}) become
\begin{equation}
\label{Piej}
\pi_{j} =\frac{p_j\e^{-rL_j/v_+}}{ \sum_{l=1}^Np_l\e^{-rL_l/v_+}},
\end{equation}
and
\begin{eqnarray}
\label{Tk2}
\pi_jT_j& =&\left \{\frac{1}{ \sum_{l=1}^Np_l\e^{-rL_l/v_+}}\right \}\left \{  \frac{p_jL_j}{v_+}\e^{-rL_j/v_+}  +\bar{p}\pi_j(\bar{\sigma}+\bar{\tau})\right . \\
&&\left .+\pi_j\sum_{l=1}^Np_l \left [\left (\bar{\tau}+ \frac{1}{r} \left (1+\frac{v_+}{v_-}\right ) \right )\left (1-\e^{-rL_l/v_+}\right )-\frac{L_l}{v_+}\left (1+\frac{v_+}{v_-}\right ) \e^{-rL_l/v_+} \right ]\right \}.\nonumber 
\end{eqnarray}
As in the single target case, $T_j\rightarrow \infty$ in the limit $r\rightarrow 0$ due to the presence of the term $\bar{\sigma}$. Finally, summing both sides of equation (\ref{Tk2}) with respect to $j$ yields the unconditional MFPT
\begin{align}
\label{Ttot}
T&=\sum_{j=1}^N\pi_jT_j=\left \{\frac{1}{ \sum_{l=1}^Np_l\e^{-rL_l/v_+}}\right \}\bigg \{
\sum_{j=1}^N \frac{p_jL_j}{v_+}\e^{-rL_j/v_+}+\bar{\tau}\left (1-\sum_{l=1}^Np_l \e^{-rL_l/v_+}\right )\nonumber \\
&\quad +\frac{1}{r} \left (1+\frac{v_+}{v_-}\right )\left [\bar{p}+\sum_{l=1}^Np_l\e^{-rL_l/v_+}\left (\e^{rL_l/v_+} -\left [1+\frac{rL_l}{v_+}\right ]\right )\right ]\bigg \}.
\end{align}

\section{Multiple search-and-capture events}

\begin{figure}[t!]
\centering
\includegraphics[width=12cm]{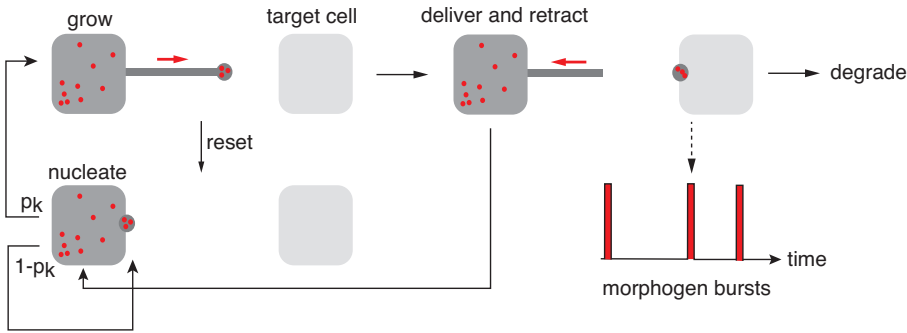} 
\caption{Multiple search-and-capture events. Alternating periods of growth, shrinkage, nucleation and target capture generates a sequence of morphogen bursts in a given target cell that is analogous to the arrival of customers in a queuing model. This results in the 
accumulation of morphogen within the cell, which is 
the analog of a queue. Degradation corresponds to exiting of customers after being serviced by an infinite number of servers.}
\label{figq}
\end{figure}

We now
consider the statistics of morphogen accumulation in the target cells in response to  multiple rounds of search-and-capture events. We assume that the build up of resources within each target is counterbalanced by degradation, so that there is a steady-state amount of morphogen in the long-time limit. The various stages are illustrated in Fig. \ref{figq}, where morphogen localized at the tip of a growing cytoneme is delivered as a ``morphogen burst'' whenever the cytoneme makes temporary contact with a target cell before subsequently retracting. 
We will assume that the total time for the particle to unload its cargo, return to the nucleation site and start a new search process is given by the random variable $\widehat{\tau}$, which for simplicity is taken to be independent of the location of the targets. (This is reasonable if the sum of the mean loading and unloading times is much larger than a typical return time.) Let $n\geq 1$ label the $n$-th capture event and denote the target that receives the $n$-th packet by $j_n$. If ${\mathbb T}_n$ is the time of the $n$-th capture event, then the inter-arrival times are 
\begin{equation}
\Delta_n:={\mathbb T}_n-{\mathbb T}_{n-1}=\widehat{\tau}_n+{\mathcal T}_{j_n},\quad n\geq 1,
\end{equation}
with $\E[{\mathcal T}_{j}]=\pi_jT_j$. Here $\pi_j$ is the splitting probability of being captured by the $j$-th target and $T_j$ is the corresponding conditional MFPT, see equations (\ref{Pie}) and (\ref{Tk}), respectively.
Finally, given an inter-arrival time $\Delta$, we denote the identity of the target that captures the particle by ${\mathcal K}(\Delta)$. We can then write for each target $j$,
\begin{eqnarray}
\label{F1}
F_{j}(t)&= \P[\Delta <t,{\mathcal K}(\Delta)=j]= \P[\Delta <t,|{\mathcal K}(\Delta)=j]\P[{\mathcal K}(\Delta)=j] \\
&=\pi_{j} \int_0^{t}{\mathcal F}_{j}(\Delta)d\Delta ,\nonumber
\end{eqnarray}
where ${\mathcal F}_{j}(\Delta)$ is the conditional inter-arrival time density for the $j$-th target. Let $\rho(\widehat{\tau})$ denote the waiting time density of the delays $\widehat{\tau}_n$. Then
\begin{eqnarray*}
 {\mathcal F}_{j}(\Delta)&=\int_0^{\Delta}dt\int_0^{\Delta} d\widehat{\tau} \delta(\Delta -t-\widehat{\tau})f_{j}(t)\rho(\widehat{\tau})=\int_0^{\Delta} f_{j}(t) \rho(\Delta-t)dt,
\end{eqnarray*}
where $f_{j}(t)$ is the conditional first passage time density for a single search-and-capture event that delivers a packet to the $j$-th target. In particular,
\begin{equation}
\label{cMFPT}
T_{j}=\int_0^{\infty} tf_{j}(t)dt.
\end{equation}
Laplace transforming the convolution equation then yields
\begin{equation}
\label{calF}
\widetilde{{\mathcal F}}_{j}(s)=\widetilde{f}_{j}(s)\widetilde{\rho}(s).
\end{equation}

As we have previously shown elsewhere \cite{Bressloff19,Bressloff20c}, the steady-state distribution of resources accumulated by the targets can be determined by reformulating the model as a G/M/$\infty$ queuing process \cite{Takacs62,Liu90}. Here we simply state the results for the steady-state mean and variance. Let $M_k$ be the steady-state number of resource packets in the $k$-th target. The mean is then
\begin{equation} \label{mean}
\overline{M}_k  =  \frac{\pi_k}{\gamma \sum_{j=1}^N \pi_{j} (T_j +\tau_{\rm cap} )}=\frac{\pi_k}{\gamma(T+\tau_{\rm cap})},
\end{equation}
where $\tau_{\rm cap}=\int_0^{\infty}\rho(\tau)d\tau$ is the mean loading/unloading time and $T=\sum_{j=1}^N\pi_jT_j$ is the unconditional MFPT (\ref{Ttot}).
Equation (\ref{mean}) is consistent with the observation that $T +\tau_{\rm cap} $ is the mean time for one successful delivery 
of a packet to any one of the targets and initiation of a new round of search-and-capture. Hence, its inverse is the mean rate of capture events and $\pi_k$ is the fraction that are delivered to the $k$-th target (over many trials). (Note that equation (\ref{mean}) is known as Little's law in the queuing theory literature \cite{Little61} and applies more generally.) The dependence of the mean $\overline{M}_k$ on the target label $k$ specifies the steady-state allocation of resources across the set of targets. Similarly, the variance of the number of resource packets is  \cite{Bressloff19,Bressloff20c}
\begin{eqnarray}
\label{var}
 \mbox{Var}[M_k]=\overline{M}_k\left [ \frac{\pi_{k}\widetilde{\mathcal F}_{k}(\gamma)}{1-\sum_{j=1}^N\pi_{j} \widetilde{\mathcal F}_{j}(\gamma)}+1-\overline{M}_k
\right ]. 
\end{eqnarray}
It can be seen that although the mean $\overline{M}_k$ only depends on the quantities $\pi_k$ and $T_k$ calculated in \S 4, higher-order moments involve the Laplace transform of the full first passage time density $f_j(t)$, see equation (\ref{calF}), which is non-trivial to calculate. Here we will focus on the mean distribution of resources, which can be interpreted as the averaged morphogen gradient. We hope to develop the analysis of higher-order moments elsewhere.

\section{Example: single-layer of target cells}

\begin{figure}[t!]
\centering
\includegraphics[width=12cm]{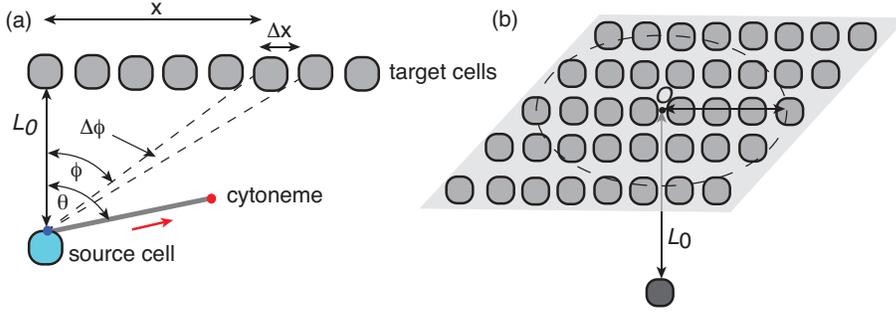} 
\caption{(a) Single 1D layer of target cells and a single source cell that extends a cytoneme at a random orientation $\theta \in [0,\pi/2]$. (b) Single 2D layer of target cells.}
\label{fig7}
\end{figure}

Consider a single layer of target cells as shown in Fig. \ref{fig7}(a). We assume that there is a single source cell that extends a cytoneme in a random direction $\theta \in [0,\pi/2]$ in the plane as illustrated in the diagram. (For the sake of illustration, we take the maximum orientation to be $\pi/2$; however, this is not a necessary condition.) If the layer subtends an angle $\Theta<\pi/2$ with respect to the source cell, then the cytoneme will extend in the wrong direction whenever $\theta> \Theta$. We also take the target cells to be sufficiently close together so that there are no ``gaps'' between them. A simple trigonemetric calculation can be used to estimate the probability that the cytoneme extends towards the $k$-th target. Suppose that the $k$-th target cell subtends an angle $\Delta \phi_k$ and that the distribution of cytoneme directions is uniform on $[0,\pi/2]$. It follows that $p_k\approx 2\Delta \phi_k/\pi$. Moreover,
$\Delta \phi_{k+1}\approx \phi(x_k+\Delta x)-\phi(x_k)$, where $\tan \phi(x)=x/L_0$, $L_0$ is the perpendicular distance of the source cell from the target layer, and $x_k=k\Delta x$. That is,
\[\Delta \phi_{k+1}\approx \mbox{sec}^{-2}(\phi_k)\frac{\Delta x}{L_0} =\left (1+\frac{x_k^2}{L_0^2}\right )^{-1}\frac{\Delta x}{L_0}.\]
Hence
\begin{equation}
\label{parray}
p_{k+1}\approx \frac{2\Delta x L_0}{\pi L_k^2},\quad L_k^2=L_0^2+x_k^2, \quad k=0,\ldots ,N-1.
\end{equation}

\begin{figure}[b!]
\centering
\includegraphics[width=12cm]{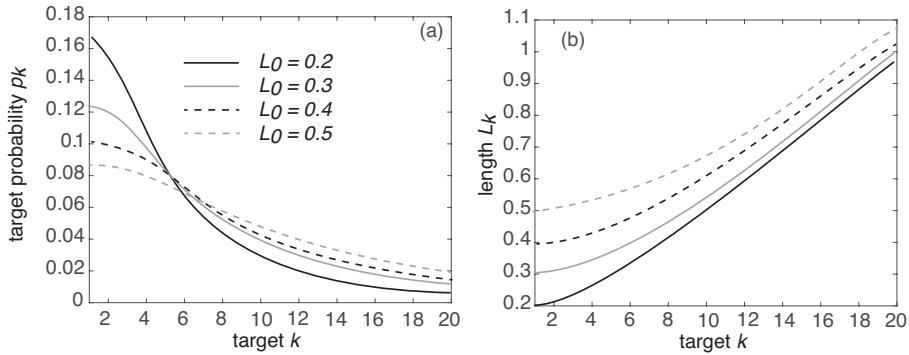} 
\caption{Plots of (a) probability $p_k$ and (b) distance $L_k$ as a function of $k$ for $N=20$ and various lengths $L_0$.}
\label{fig8}
\end{figure}

In the following we will fix the length scale by taking the total size of the target array to be $L_{\rm target}=1$ and set $\Delta x=L_{\rm target}/N$. Similarly, the time scale is fixed by setting $v_+=1$. Experimental studies of cytoneme-mediated transport of Wnt morphogen in zebrafish \cite{Stanganello16} and Shh in chicken \cite{Sanders13} indicate that the growth rate of a cytoneme is of the order $v_+\sim 0.1\  \mu$m/s. Cytoneme lengths vary from $10-100\ \mu$m so if we take $L_{\rm target}=1$ to correspond to a length of 25 $\mu$m (around 20 cells), then the fundamental time-scale is 250 s (around 4 minutes). We will mainly focus on the dependence of the morphogen distribution on $L_0,N,r$ by taking $v_-\gg v_+$ (fast return speed). Finally, a new cytoneme can be formed approximately twice every minute. Therefore, we take $\tau_{\rm ref}=\tau_{\rm cap}=0.1$ and assume that degradation occurs on a time-scale of hours by setting $\gamma^{-1}=100$.  

\begin{figure}[t!]
\centering
\includegraphics[width=12cm]{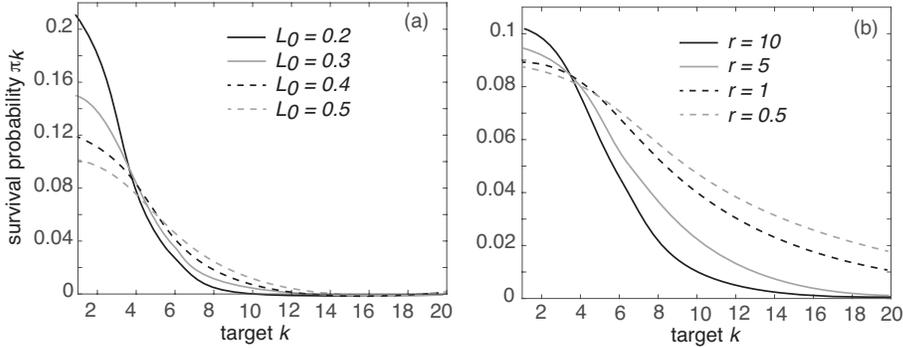} 
\caption{Plots of splitting probability $\pi_k$ as a function of $k$ for $N=20$: (a) various $L_0$ and $r=1$; (b) various $L_0$ and $r=10$; (c) various resetting rates $r$ and $L_0=1$.}
\label{fig9}
\end{figure}

\begin{figure}[b!]
\centering
\includegraphics[width=6cm]{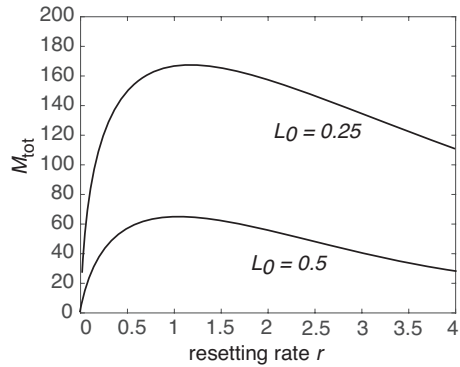} 
\caption{Steady-state mean number of resources $M_{\rm tot}$ delivered to all of the targets as a function of the resetting rate $r$ for $L_0=0.25,0.5$. Other parameters are $\tau_{\rm ref}=\tau_{\rm cap}=0.1$, $\gamma=0.01$, $v_+=1$ and $v_-=100v_+$}
\label{fig10}
\end{figure}

In Fig. \ref{fig8} we plot the target probability $p_k$ and target-source separation $L_k$ as a function of $k$ for various choices of $L_0$ and $N=20$. As expected, reducing the perpendicular distance of the source cell from a target array of fixed length leads to a greater variation of $p_k$ and $L_k$ with $k$. In addition, the total probability $p_{\rm tot}=\sum_{k=1}^Np_k$ increases as $L_0$ decreases. In Fig. \ref{fig9} we show plots of the splitting probability $\pi_k$ as a function of $k$, which is obtained by substituting equation (\ref{parray}) into (\ref{Piej}). It can be seen that reducing $L_0$ for fixed resetting rate sharpens the spatial ($k$-dependent) variation of $\pi_k$ along the target array, as does increasing $r$ for fixed $L_0$. Note in particular that fast resetting significantly amplifies the spatial variation of $\pi_k$ compared to $p_k$, see Figs. \ref{fig9}(a) and \ref{fig8}(a).

It follows from equation (\ref{mean}) that $\pi_k$ determines the corresponding steady-state mean distribution of resources $\overline{M}_k$ up to the normalization factor 
\begin{equation}
\overline{M}_{\rm tot}:=\sum_{k=1}^N\overline{M}_k=\frac{1}{\gamma(T+\tau_{\rm cap})},
\end{equation}
where $T$ is the unconditional MFPT. The latter is determined by substituting equation (\ref{parray}) into (\ref{Ttot}). Hence, while $\pi_k$ specifies the relative distribution of resources to the target cells, that is the shape and steepness of the morphogen gradient, $T$ fixes the total amount delivered to all the target cells. As in the case of a single target, see Fig. \ref{fig5}, $T$ has a minimum at an optimal resetting rate $r_{\rm tot}$, which implies that $M_{\rm tot}$ has a maximum at the same value of $r$. This is illustrated in Fig. \ref{fig10}. Our analysis suggests that although varying the rate of resetting controls the steepness of the gradient, $r$ should lie in an interval around $r_{\rm opt}$ in order to ensure sufficient resources are delivered to the targets. 
Finally, note that a similar construction can be applied to a 2D layer of target cells as illustrated in Fig. \ref{fig10}(b). To a first approximation, the probability $p_k$ and distance $L_k$ of the $k$-th target cell will depend on the in-plane radial distance of the target cell from the point $O$ where the perpendicular projection from the source cell intersects the layer.

\section{Discussion}

In this paper we generalized our previous search-and-capture model of cytoneme-based morphogenesis \cite{Bressloff19,Bressloff20c} in order to take into account the direction of cytoneme growth. Assuming that each time the cytoneme nucleates from a source cell it grows in a random direction means that a single source cell can explore a two- or three-dimensional domain containing a distribution of target cells. However, it also implies that the cytoneme can grow in a direction that misses any target cell (search failure). This would lead to unrealistically large search times in the absence of a resetting mechanism that allows the cytoneme to retract and nucleate in a different direction. Using renewal theory we showed that the distribution of morphogen (morphogen gradient) generated by a single source cell is determined by the splitting probabilities $\pi_k$ for the cytoneme capture to be captured by the $k$-th target cell, $k=1,\ldots,N$ and the unconditional MFPT for target capture. The splitting probability $\pi_k$ depends on the probability $p_k$ that the cytoneme is oriented towards the given cell, which itself depends on the distance $L_k$ of the target cell from the source cell and its size. 

\begin{figure}[t!]
\centering
\includegraphics[width=12cm]{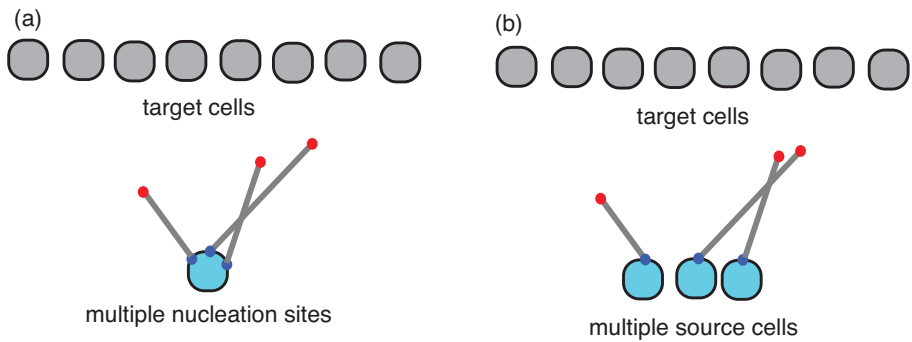} 
\caption{Parallel search-and-capture events due to (a) multiple nucleation sites and (b) multiple source cells.}
\label{fig11}
\end{figure}

One obvious extension of the current theory is to consider a population of source cells or a single source cell with multiple nucleation sites, see Fig. \ref{fig11}, or a combination of the two. Multiple source cells clearly complicates the analysis because one now has to take into account the different positions of the source cells relative to the population of target cells. In other words, the probabilities $p_k$ for each source cell would differ. For simplicity, suppose that the source cells are clustered in a sufficiently localized spatial region so that the $p_k$ are taken to be source-independent, and assume that the nucleation of each cytoneme occurs independently. In that case, one can directly carry over the analysis of this paper. In particular, suppose that there is a total of $N_c$ independently nucleating cytonemes with the same mean nucleation time.
 Let $M_k^{\mu}(t)$ be the number of morphogens present in the $k$-th target cell at time $t$ that were delivered by the $\mu$-th cytoneme and set
$M_k^\Sigma(t) = \sum_{\mu=1}^{\mathcal M} M_k^{\mu}(t)$.
Since the $M_k^{\mu}(t)$ are independent identically distributed random variables, we have the steady-state mean and variance
\begin{equation*}
	\langle M_k^\Sigma \rangle = N_c \langle M_k \rangle,\quad \mbox{Var}[M_k^\Sigma] =N_c \mbox{Var}[M_k].
\end{equation*}
Thus multiple independent cytonemes scale up the morphogen concentration gradient by a factor of $N_c$, and reduce fluctuations according to $N_c$.

There are a number of other possible details that could be included in future work. First, we treated each cytoneme as a rigid filament, whereas it is likely that they are semi-flexible, which means that the orientation of the morphogen tip could change as the cytoneme grows. Second, in certain developmental systems, one finds that both target and source cells extend cytonemes that make contact at intermediate locations between the cells. Third, in systems such as the neural plate of zebrafish \cite{Stanganello16}, morphogenesis occurs in a growing tissue domain. The directional search-and-capture model presented in this paper provides a mathematical framework for investigating the role of these various factors as well as tissue geometry on the formation of cytoneme-based morphogen gradients.

\begin{footnotesize}

\end{footnotesize}

\end{document}